\documentclass[twocolumns,bibyear]{aa} 
\usepackage{graphicx,epstopdf}
\usepackage{url}
\usepackage{txfonts}
\usepackage{amsmath}
\usepackage{chngcntr}
\counterwithout{figure}{section}
\def\Q{\ifmmode\mathcal{Q}\else$\mathcal{Q}$\fi}

\begin{document} 

   \title{The galaxy-wide stellar initial mass function in the presence of cluster-to-cluster IMF variations\thanks{The grid of IGIMF models is available at the CDS via anonymous ftp to cdsarc.u-strasbg.fr (130.79.128.5)
or via http://cdsweb.u-strasbg.fr/cgi-bin/qcat?J/A+A/}}

   \author{Sami Dib\inst{1}}
 
   \institute{Max Planck Institute for Astronomy, K\"{o}nigstuhl 17, 69117, Heidelberg, Germany\\
      \email{sami.dib@gmail.com; dib@mpia.de}
                      }
          
\authorrunning{Dib}
\titlerunning{The IGIMF with a variable IMF in clusters}
         
 
\abstract{
We calculate the stellar integrated galactic initial mass function (IGIMF) in the presence of cluster-to-cluster variations of the IMF. Variations of the IMF for a population of coeval clusters that populate the initial cluster mass function (ICLMF) are taken into account in the form of Gaussian distribution functions of the IMF parameters. For the tapered power-law function used in this work, these are the slope at the high-mass end, $\Gamma$, the slope at the low-mass end, $\gamma$, and the characteristic mass $M_{ch}$. The level of variations is modeled by varying the width of  the Gaussian distributions. The reference values are the standard deviations of the parameters observed for the population of young clusters in the present-day Milky Way, which are $\sigma_{\Gamma}=0.6$, $\sigma_{\gamma}=0.25$, and $\sigma_{M_{ch}}=0.27$ M$_{\odot}$. We find that increasing the levels of dispersion for $\gamma$ and $\Gamma$ tends to moderately flatten the IGIMF at the low and high-mass end, respectively. The characteristic mass of the IGIMF is, however, strongly impacted by variations in $M_{ch}$. Increasing the value of $\sigma_{M_{ch}}$ shifts the peak of the IGIMF to lower masses, rendering the IGIMF more bottom heavy. This can provide a simple explanation for the bottom-heavy stellar mass function that is inferred for early-type galaxies since these are likely the result of a merger of disk galaxies where the physical conditions of the star-forming gas may vary significantly both in time and space in the merging system. The effect of IMF variations on the IGIMF is compared to the effects of other processes and sources of systematic variations such as those due to variations in the shape of ICLMF, the gas-phase metallicity, and the galactic star formation rate (SFR) which can potentially affect the maximum mass of stellar clusters in a galaxy and set the mean value of the characteristic mass in clusters. For the various dependencies we have explored, we found that the effect of IMF variations is a dominant factor that always affects the characteristic mass of the IGIMF. For the regimes at low metallicity where the IGIMF resembles a single power law, an increased level of IMF variations renders the IGIMF steeper and more bottom heavy, especially at low SFRs. On the other hand, variations in the IMF in the high mass regime can be easily dominated by variations in the slope of the ICLMF. We compare our results of the metallicity and SFR-dependent IGIMF to a sample of Milky Way ultra-faint dwarf (UFD) satellite galaxies that have available metallicity measurements. The present-day stellar mass function of these galaxies is a good analog to the IGIMF at the time their overall population of stars formed. We show that the slope of the stellar mass function of the UFD galaxies measured for stars in the mass range [0.4,0.8] M$_{\odot}$ can only be reproduced when IMF variations of the same order as those measured in the present-day Milky Way are included. Our results suggest that the inclusion of IMF variations in models of galaxy formation and evolution is of vital importance in order to improve our understanding of star formation and star formation feedback effects on galactic scales.       
}


 \maketitle

%

\section{Introduction}\label{introduction}

The initial mass function (IMF) of stars (i.e., the distribution of the masses of stars at their birth) is of fundamental importance for many branches in astrophysics. The IMF holds information on the physics of the star formation process and on its potential environmental dependencies. Its shape affects the efficiency of star formation in molecular clouds (e.g., Dib 2011; Dib et al. 2011,2013; Hony et al. 2015), and constrains the amount of feedback from stars into the large-scale interstellar medium (e.g., Dib et al. 2006;2009; Haid et al. 2016; Gatto et al. 2017; Dib et al. 2021). Furthermore, the IMF determines the chemical and dynamical evolution of galaxies (e.g., Larson 1998; Bekki 2013; Dav\'{e} et al. 2011a,b; Brusadin et al. 2013; Few et al. 2014; Recchi \& Kroupa 2015; Matteucci 2016; Vincenzo et al. 2016; Fontanot et al. 2017). Intense efforts have been devoted to the characterization of the shape of the mass function of stars, both on cluster and galactic scales, in the Milky Way and in external galaxies. For nearby stars in the Galactic field, Salpeter (1955) found that the stellar mass function is well described by a power law $\left(dN/d{\rm log}M_{*}\right)=M_{*}^{-\Gamma}$, where $dN$ is the number of stars between ${\rm log}M_{*}$ and ${\rm log} M_{*}+d{\rm log}M_{*}$, with $\Gamma \approx 1.35$. Subsequent surveys of larger volumes of the Milky Way have enabled a more complete characterization of the shape of the present-day stellar mass function down to the low-mass and substellar mass regimes. The results from these surveys suggest that the mass function of stars in the Galactic field, which has not been corrected for the effects of the binary population, rises from the brown dwarf and low stellar mass regime until it peaks at $\approx 0.25-0.4$ M$_{\odot}$ after which point it declines steeply in the intermediate-to-high mass regime (e.g., Miller \& Scalo 1979; Scalo 1986; Bochanski et al. 2010; Parravano et al. 2011). In a more recent determination of the Galactic IMF, Mor et al. (2019) used constraints from both the chemical composition of stars (APOGEE data) and from their dynamics and spatial distribution (GAIA data) and, most importantly, they relaxed the assumption of a constant star formation rate (SFR) made in earlier studies. These authors found that the time-averaged Galactic IMF is shallower at both the low- and high-mass ends than what was previously assumed. This result has been predicted by Dib \& Basu (2018) who attribute it to the existence of cluster-to-cluster variations of the IMF in Galactic clusters.

For individual stellar clusters, there is a large body of estimates of the IMF, both in the Milky Way and in external nearby galaxies with discrepant conclusions in terms of their similarity with the Galactic field stellar mass function (Bastian et al. 2010; Kroupa et al. 2013; Dib 2014; Dib et al. 2017). Some of the earlier studies argued that the observed level of variations between the IMFs of Galactic clusters is compatible with the level of scatter that can be expected from stochastic sampling in clusters of a given mass (Kroupa 2002). However, recent studies suggest that there are non-negligible cluster-to-cluster variations in the set of parameters that characterize the shape of the IMF among the population of young Galactic stellar clusters (e.g., Sharma et al. 2008; Massey 2011; Dib 2014; Mallick et al. 2014; Tripathi et al. 2014; Kraus et al. 2017; Dib et al. 2017; Zhang et al. 2018; Schneider et al. 2018; Kalari et al. 2018; Dib et al. 2022). In particular, Dib et al. (2017) show that the fraction of single O stars measured in a large sample of Galactic clusters (341 clusters from the Milky Way Stellar Clusters Survey; Kharchenko et al. 2013) can only be reproduced by populations of Galactic clusters that have a significant intrinsic scatter in the set of parameters that characterize their IMFs. In M31, the comparison of the inferred slopes of 85 clusters by Weisz et al. (2015) also suggests the existence of non-negligible variations of the IMF in the intermediate- to high-mass regime (see Figure 4 in their paper, but also see their section 3.2 for a somewhat different result and conclusion).   

Understanding whether the stellar mass function is universal on galactic scales has also received significant attention, both in early- and in late-type galaxies (e.g., La Barbera et al. 2016; Vaughan et al. 2016; van Dokkum et al. 2017a; Sarzi et al. 2017). Assessing the shape of the IMF can be made by matching the mass-to-light ratio ($M/L$) of galaxies which is derived from dynamical modeling to the ($M/L$), which is derived from fitting the spectrum of the galaxy with population synthesis models (e.g., Dutton et al. 2012; Capellari et al. 2012; Tortora et al. 2012; Conroy et al. 2013; Lyubenova et al. 2016). Other methods rely on measuring the spectral features that are sensitive to the population of low-mass and dwarf stars (e.g., Cennaro et al. 2003; van Dokkum \& Conroy 2010; Smith et al. 2012; Ferreras et al. 2013; Spiniello et al. 2014; McConnell et al. 2016; Parikh et al. 2018), and on matching the galactic stellar mass derived using both stellar population synthesis models and gravitational lensing (e.g., Ferreras et al. 2008; Auger et al. 2010; Treu et al. 2010; Posacki et al. 2015; Leier et al. 2016). Most of these studies were conducted for early-type galaxies and the emerging picture is that the stellar mass function in these galaxies is bottom heavy (i.e., has a higher fraction of low-mass stars than for a Milky Way-like stellar mass function). These studies also find that the slope of the stellar mass function slope in the intermediate- to high-mass end is correlated with the stellar velocity dispersion in that a steeper slope is measured for galaxies with a higher velocity dispersion (Conroy et al. 2013; Capellari 2016). Similar galaxy-to-galaxy variations with similar dependencies on galactic properties have been corroborated for early-type galaxies at higher redshift (e.g., Shetty \& Capellari 2014,2015; Gargiulo et al. 2015). Galaxy-to-galaxy variations as well as radial variations of the galactic stellar mass function have also been reported for dwarf and spiral galaxies. A steeper slope with increasing metallicity has been reported in disk galaxies by Mart\'{i}n-Navarro et al. (2015). Geha et al. (2013) and Gennaro et al. (2018) have reported shallower than Salpeter slopes in UFD galaxies in the intermediate stellar mass regime (in the stellar mass range $\approx 0.4$ to $\approx 0.8$ M$_{\odot}$). Brewer et al. (2012) show that lens models with a Salpeter mass function are unable to fit the observations for the low-mass spiral galaxies in their sample. Zhang et al. (2007) found that the slope of the galactic stellar mass function in disk galaxies is steeper at higher metallicities. Li et al. (2017) and Bernardi et al. (2018) show that the mass function of disk galaxies with higher stellar velocity dispersion tends to be more bottom heavy, in a similar fashion to what is found in early-type galaxies. In contrast, a top-heavy stellar mass function in starbursts has been proposed as an explanation for the high mass-to-light ratios in ultra-compact dwarf galaxies (Dabringhausen et al. 2009). Using a sample of $\approx 33000$ galaxies, Gunnawardhana et al. (2011) argued for the existence of an SFR-mass function relation, such that galaxies with a higher SFR have a more top-heavy stellar mass function. A similar argument has been made about the mass function of stars in the accretion disk in the center of the Galaxy (e.g., Nayakshin \& Sunyaev 2005) as well as in starburst clusters (Stolte et al. 2005; Schneider et al. 2018, Hosek et al. 2019). There are, however, cautionary arguments that suggest that a better assessment of the role of the different uncertainties is needed before a statement on galaxy-to-galaxy variations of the stellar mass function can be made (e.g., Tang \& Worthey 2015; Clauwens et al. 2015; Newman et al. 2017).

In this work, we measure the effect of cluster-to-cluster variations of the IMF on the integrated galactic initial mass function (IGIMF). We also compare the effect of cluster-to-cluster variations of the IMF to variations in other quantities that can modify the shape of the IGIMF such as variations in the galactic SFR, metallicity, and variations of the mean characteristic mass of the IMF. The main difference to existing models of the IGIMF developed by other authors is that we allow for cluster-to-cluster variations of the parameters of the IMF for any population of clusters that form in the galaxy. In Sect.~\ref{previous} we briefly summarize previous work on the IGIMF. The IMF model used in this work is presented in Sect.~\ref{imf}, along with a discussion on the level of variations of the parameters of the IMF measured in the Milky Way and in nearby galaxies as well as on their possible physical origin. In Sect.~\ref{iclmf}, we briefly recall the shape of the ICLMF and discuss its variations. In Sect.~\ref{igimf}, we present our results and this is followed by a comparison to observations in Sect.~\ref{compobs}. In Sect.~\ref{conc}, we conclude.

\section{The concept of the IGIMF and brief summary of previous work }\label{previous}

Under the assumption that all (or most) stars form in clusters and associations (an assumption supported by the results of Dinnbier et al. 2022), the IGIMF is the sum of all the IMFs of its star cluster constituents. The IGIMF is the total galactic population for a single generation of stars (i.e., stars born within a short lapse of time compared to the galaxy's ages) and for a fixed SFR. Models of the IGIMF were presented by Kroupa \& Weidner (2003) and Weidner \& Kroupa (2005) and used in several other works to explain a number of galactic properties (e.g., Weidner et al. 2011; Haas \& Anders 2011; Weidner et al. 2013a,b; Yan et al. 2020). Usually, models of the IGIMF assume a correlation between the maximum mass of a cluster that can form in the galaxy and the galactic SFR which is derived from observations (e.g., Weidner et al. 2004; Gonz\'{a}lez-L\'{o}pezlira et al. 2012) and they can also take into account that the shape of the Initial CLuster Mass Function (ICLMF) depend on the galactic SFR. In some models, given a cluster mass $M_{cl}$, a maximum stellar mass ($M_{*,max}$)-cluster mass relation is imposed for each individual cluster that forms within the galaxy. However, all previous models have assumed that the IMF of each individual cluster is indistinguishable from that of the mass function of the Galactic field, aside from its normalization which is determined by the available stellar mass in each cluster. Later versions of the IGIMF models adopted a metallicity and density dependence of the slope at the high-mass end which is based on available observational constraints (Marks et al. 2012). Using these prescriptions in full, or partially, a galactic IMF can be calculated either for the galaxy as a whole or for local regions within a galaxy taking into account local or radial variations of the metallicity (e.g., Pflamm-Altenburg \& Kroupa 2008, Yan et al. 2017; Je\v{r}\'{a}bkov\'{a} et al. 2018).

A variant of the IGIMF theory has been presented by Dib \& Basu (2018) in which variations of the IMF in clusters were incorporated into the derivation of the IGIMF, but without imposing some of the other constraints or assumptions such as the ${\rm{SFR}}-M_{cl,max}$ and the $M_{*,max}-M_{cl}$ relations. Dib \& Basu (2018) show that a Galactic field-like mass function can result from the summation of a population of clusters with intrinsic cluster-to-cluster variations, even when the mean values of the IMF parameters of the clusters are different from the ones of the Galactic field. They also found that the IGIMF in the presence of cluster-to-cluster IMF variations is shallower at both the high- and low-mass end, owing to the contribution from the tails of the distributions of the parameters that characterize the shape of the IMF in those stellar mass regimes. In contrast, Je\v{r}\'{a}bkov\'{a} et al. (2018) constructed a grid of IGIMF models in which they fixed the IMF of clusters to a Kroupa- like IMF (Kroupa 2002) and considered various forms of the ICLMF, the SFR of the galaxy, and included a metallicity and density dependence for the IMF of clusters. 

Deriving the shape of the IGIMF and exploring its physical dependencies is important for understanding the origin of the observed present-day mass function in galaxies. The latter can be obtained by integrating the IGIMF over cosmic time. In doing so, one has to take into account, in addition to the role of the time-varying SFR and potential time variations of the IMF and of the ICLMF (and of its cutoffs), the effects of stellar evolution such as the effects of mass loss by stellar winds, the removal of massive stars that turn into compact objects, and the effects of an evolving gas-phase metallicity in the galactic disk out of which new generations of star will form. Fontanot (2014) present a model that tracks the evolution of metals with simplifying assumptions about the metal enrichment (i.e., instantaneous recycling of metals) and tested various forms of the IMF.  He found that a top-heavy galactic mass function (with respect to a Kroupa Milky Way-like mass function) can be produced in high SFR environments. Hasani Zoonoozi et al. (2019) present a similar model to reproduce the present-day stellar mass function of the Milky Way as well as other Galactic observables such as the total Galactic stellar mass and mass-to-light ratio. As stressed above, the only model that has considered the effects of a nonuniversal IMF in clusters that formed at a given cosmic epoch is the one presented by Dib \& Basu (2018). In this work, we extend this model of the IGIMF to cases where we impose different additional constraints, not considered in Dib \& Basu (2018), such as the ${\rm SFR}-M_{cl,max}$ relation as well as a metallicity dependence. Below, we describe the different elements of the model. 

\section{The IMF of individual clusters}\label{imf}

\subsection{Mathematical representation of the IMF}\label{mathimf}

We adopt a description of the IMF in clusters that is given by the tapered power-law (TPL) function (de Marchi et al. 2010; Parravano et al. 2011; Dib et al. 2017). The TPL function is a convenient form to describe the IMF since it possesses only three free parameters. We also adopt this functional form because Dib et al. (2017) inferred the distribution functions of the three parameters of the TPL function for young Galactic clusters. The TPL function is given by

\begin{equation}
\phi\left(M_{*}\right)=\frac{dN}{dM_{*}}=A_{*}\times M_{*}^{-\Gamma-1}\left\{1-\exp\left[-\left(\frac{M_{*}}{M_{ch}}\right)^{\gamma+\Gamma}\right] \right\},
\label{eq1}
\end{equation} 

\noindent where $\Gamma$ is the slope in the intermediate to high stellar mass regime, $\gamma$ is the slope in the low-mass regime, $M_{ch}$  is the characteristic mass, and $A_{*}$ is the normalization coefficient which is set by the cluster's mass such that $M_{cl}=A_{*}\int_{M_{*,min}}^{M_{*,max}} \phi({\rm log}M_{*}) dM_{*}$, with $M_{*,min}$ and $M_{*,max}$ being the minimum and maximum stellar masses, respectively. Using the TPL function, it is straightforward to include additional empirical dependencies  informed from the observations. Such dependencies could include a dependence on the gas-phase metallicity and mean gas density of the star-forming gas. Marks et al. (2012) propose that the dependence of the slope at the high-mass end (in the $dN/dM \propto M^{-\alpha}$ formulation) on metallicity ([Fe/H]) and total density of the protocluster cloud (gas+stars, $\rho_{pcl}$) is given by 

\begin{equation}
\alpha  = \left\{ 
\begin{array}{ll} 
2.3                 \hspace{2.15cm} \mathrm{if}\, x< -0.87\,,\\
-0.41x+1.94 \,\hspace{0.5cm}   \mathrm{if}\, x\geq-0.87\,, 
\end{array} \right.
\label{eq2} 
\end{equation} 

\noindent where $\alpha=\Gamma+1$, and $x$ is given by 

\begin{equation}
x=-0.14\left[{\rm Fe/H} \right]+0.99 {\rm log_{10}}\left(\frac{\rho_{pcl}}{10^{6}{\rm M}_{\odot} {\rm pc}^{-3} }  \right).
\label{eq3}
\end{equation}

In what follows, we calculate models of the IGIMF with and without using the constraint coming from Eq.~\ref{eq2}. It is useful to point out that in the data presented by Marks et al. (2012), the dependence of the high-mass slope of the IMF on metallicity is mostly driven by a few points pertaining to low metallicity globular clusters $({\rm Fe/H}] \lesssim -2.$, see Figure 4 in their paper).

\subsection{Mean values and scatter of the IMF parameters}\label{meanparam}

Measurements of the parameters that describe the shape of the IMF ($\Gamma$, $\gamma$, and $M_{ch}$ for the TPL function) display a certain amount of scatter. Dib (2014) show that the mean values of $\Gamma$, $\gamma$, and $M_{ch}$ for the star system IMFs (i.e., uncorrected for binarity) for a sample of eight young Galactic clusters is $\Gamma_{obs}=1.37$, $\gamma_{obs}=0.91$, and $M_{ch,obs}=0.41$ M$_{\odot}$ with standard deviations of $\sigma_{\Gamma_{obs}}=0.6$ ($\approx 0.45$ dex), $\sigma_{\gamma_{obs}}=0.25$ ($\approx 0.3$ dex), and $\sigma_{M_{ch,obs}}=0.27$ M$_{\odot}$ ($\approx 0.65$ dex). Dib et al. (2017) show that such a level of intrinsic scatter around the same mean values for each of these parameters is necessary in order to reproduce the fraction of observed single O stars that is measured in young (ages $\lesssim 12$ Myrs) clusters in the Milky Way (341 clusters in the MWSC survey). Dib et al. (2017) also show that stochastic sampling effects around a fixed universal IMF do not produce enough scatter which can help explain the observed fraction of single O stars. Thus, a significant part of the scatter is expected to be of physical origin and associated with variations in the physical conditions (i.e., temperature, density, characteristics of turbulent motions, strength of the magnetic field, rate of interactions, and mergers of protostellar cores) of the clusters progenitor clouds. Variations of mean temperatures, densities, and virial parameters have been reported in several studies on molecular clouds and clumps of the Milky Way (e.g., Jackson et al. 2006; Saito et al. 2007; Dib et al. 2012; Mills \& Norris 2013; Ott et al. 2014; S\'{a}nchez-Monge et al. 2014; Svoboda et al. 2016; Urquhart et al. 2018).

The mean values and standard deviations of the parameters inferred by Dib (2014) and Dib et al. (2017) are characteristic of the present-day star formation in the Milky Way. In principle, there is no reason to believe that these values are necessarily representative of star formation in other galactic environments or of star formation at higher redshift. In terms of the mean values, the results of van Dokkum (2008) suggest a higher characteristic mass (on the order of 2 M$_{\odot}$) at redshift $\approx 4$ in massive spiral galaxies, which is larger than the mean value of $M_{ch}$ observed for young clusters in the present-day Milky Way. Dav\'{e} (2008) and Narayanan \& Dav\'{e} (2012,2013) considered the effect of a time-varying Jeans mass in molecular clouds on the buildup of the stellar population in galaxies. They found that the Jeans mass of the clouds -- which scales as $T^{3/2} n^{-1/2}$, where $T$ and $n$ are the clouds' temperature and number density, respectively -- scales with the SFR $M_{J} \propto {\rm SFR}^{k}$ M$_{\odot}$, with $k \approx 0.3-0.4$. This dependence is due to the gas-dust coupling at high densities and the direct dependence of the cosmic ray fluxes with the SFR at lower densities (see also Klessen et al. 2007). They further posit that there is a direct relationship between the characteristic mass of the IMF, $M_{ch}$, and the clouds' Jeans mass. Thus, variations in the peak of the galactic mass function can be directly related to the time-varying SFR. Narayanan et al. (2013) assume a broken power-law shape of the IMF and calibrated the value of the break point (i.e., the characteristic mass in their model) using an assumed value for the IMF of 0.5 M$_{\odot}$ (Kroupa 2002) and the present-day Galactic ${\rm SFR}$ (${\rm SFR_{\rm MW}}$), which they assume to be $2$ M$_{\odot}$ yr$^{-1}$. In this work we assume that SFR$_{\rm MW}=1$ M$_{\odot}$ yr$^{-1}$ (Robitaille et al. 2010) and that the characteristic mass is 0.42 M$_{\odot}$ (Parravano et al. 2011). With these elements, the dependence of $M_{ch}$ on the SFR can be written as follows:
 
 \begin{equation}
 M_{ch} =0.42 \times \left(\frac{\rm SFR}{\rm SFR_{MW}}\right)^{k} {\rm M}_{\odot} 
 \label{eq4} 
 ,\end{equation}
 
where $k=0.4$. Higher Jeans masses in the clouds can be reached by increasing their temperature via intense cosmic-ray heating (e.g., Hocuk et al. 2010,2011; Papadopoulos 2010; Papadopoulos et al. 2011; Fontanot et al. 2018a). Fontanot et al. (2018b) found that the cosmic-ray heating of molecular clouds can help generate IGIMF shapes which are simultaneously shallower at the high-mass end and steeper at the low-mass end than those of a Kroupa-like IMF, and they argue that such solutions can explain both the levels of $\alpha$-elements' enrichment and the excess of low-mass stars as a function of stellar mass that are observed for local spheroidal galaxies. In terms of scatter around the mean values, larger variations in the physical conditions of star-forming molecular clouds have been reported in Local Group galaxies (e.g., Colombo et al. 2015; Tang et al. 2017; Gorski et al. 2017) as well as variations between galaxies (Hughes et al. 2013; Bolatto et al. 2018). Similar large variations in the physical conditions of the star-forming gas has been observed in galaxies at high redshift and in a merging starburst system (Ott et al. 2011; Gonz\'{a}lez-Alfonso et al 2012; Mangum et al. 2013; Miyamoto et al. 2015; Zschaechner et al. 2016). For the present-day Milky Way, a level of dispersion in the characteristic mass, such as the one reported by Dib (2014) and Dib et al. (2017), is plausible and can be attributed to variations in the level of turbulent support in star-forming molecular clouds (Haugb{\o}lle et al. 2018). Other studies have suggested that variations in all of the IMF parameters can be caused by cloud-to-cloud variations in the mean level of accretion rates onto protostellar cores (Basu \& Jones 2004, Myers 2009; Dib et al. 2010), or by the coalescence of protostars in high-density environments (Elmegreen \& Shadmehri 2003; Shadmehri 2004; Dib 2007; Dib et al. 2007,2008a,b).

We describe the scatter in the IMF parameters using Gaussian distribution functions. For $\Gamma$, $\gamma$, and $M_{ch}$, they are given by  

\begin{equation}
P\left(\Gamma \right)=\frac{1}{\sigma_{\Gamma} \sqrt{2 \pi}} \exp\left(-\frac{1}{2} \left(\frac{\Gamma-\bar{\Gamma}}{\sigma_{\Gamma}}\right)^{2}\right),
\label{eq5}
\end{equation}

\begin{equation}
P\left(\gamma \right)=\frac{1}{\sigma_{\gamma} \sqrt{2 \pi}} \exp\left(-\frac{1}{2} \left(\frac{\gamma-\bar{\gamma}}{\sigma_{\gamma}}\right)^{2}\right),
\label{eq6}
\end{equation}

\begin{equation}
P\left(M_{ch} \right)=\frac{1}{\sigma_{M_{ch}} \sqrt{2 \pi}} \exp\left(-\frac{1}{2} \left(\frac{M_{ch}-\bar{M_{ch}}}{\sigma_{M_{ch}}}\right)^{2}\right).
\label{eq7}
\end{equation}

For the slopes of the IMF at the low- and high-mass end, we assume that their mean values are similar to those of the Milky Way Galactic field, namely $\bar{\Gamma}=1.35$, $\bar{\gamma}=0.58$. For the mean value of $M_{ch}$ ($\bar{M_{ch}}$), we adopt the value given by Eq.~\ref{eq4}. The standard deviations of the three parameters are parametrized as $\sigma_{\Gamma}=a_{\Gamma} \sigma_{\Gamma_{obs}}$, $\sigma_{\gamma}=a_{\gamma} \sigma_{\gamma,obs}$, and $\sigma_{M_{ch,obs}}=a_{M_{ch}} \sigma_{M_{ch,obs}}$, where $\sigma_{\Gamma_{obs}}$, $\sigma_{\gamma,obs}$, and $\sigma_{M_{ch,obs}}$ are the values measured by Dib et al. (2014) and Dib et al. (2017) which we recall are $\sigma_{\Gamma}=0.6$, $\sigma_{\gamma}=0.25$, $\sigma_{M_{ch}}=0.27$ M$_{\odot}$, and where $a_{\Gamma}$, $a_{\gamma}$, and $a_{M_{ch}}$ are free parameters. Hence, values of $a_{\Gamma}=1$, $a_{\gamma}=1$, and $a_{M_{ch}}=1$ refer to standard deviations of the parameters similar to those found in the observations. We note that Mor et al. (2019) found that values of $a_{\Gamma}=a_{\gamma}=a_{M_{ch}}=0.5$ better fit the time-averaged IGIMF they derived for the Milky Way. In the rest of work, we always vary the three parameters in unison (i.e., $[a_{\Gamma},a_{\gamma},a_{M_{ch}}]=[1,1,1]$ which for simplicity we write as $a=1$). A value of $a=0.5$ corresponds to standard deviations of the three Gaussian distributions that are half the values measured by Dib (2014) and Dib et al. (2017). 

The cutoffs of the parameters, namely [$\Gamma_{min}, \Gamma_{max}$], [$\gamma_{min},\gamma_{max}$], and $[M_{ch,min}-M_{ch,max}]$, could also vary as a function of the Galactic SFR. However, at the moment, there are no constraints as to how these quantities might depend on the SFR and how they evolve over cosmic time. In this work, we adopt a conservative approach and assume that the limits on each of these parameters are independent of the SFR. We adopt the same values used in Dib et al. (2017), namely $[M_{ch,min},M_{ch,max}]=[0.05,1]$ M$_{\odot}$, $[\Gamma_{min}, \Gamma_{max}]=[0.7,2.4]$, and $[\gamma_{min},\gamma_{min}]=[0.4,1.5]$. In cases where we change the mean values of the three parameters, as is the case when the mean value of $M_{ch}$ changes with the SFR, we also vary the standard deviations and the cutoff values in the same percentage around the mean values using the same rate of variations as the one measured by Dib et al. (2017) for the young clusters of the Milky Way. 

\section{The initial cluster mass function}\label{iclmf}

As stated in Sect.~\ref{previous}, the ICLMF is the distribution function of young, embedded, or partially embedded stellar clusters. Young clusters (with ages $\lesssim$ 10 Myrs) are those that formed most of their stellar population, but have not lost them due to internal dynamical interactions or to the effects associated with stellar evolution. A common representation of the ICLMF is a power-law function (but see also Gieles et al. 2006; Lieberz \& Kroupa 2017) given by

\begin{equation}
\xi_{cl}(M_{cl})=\frac{dN}{dM_{cl}} \propto  M_{cl}^{-\beta}.
\label{eq8}
\end{equation}

Measurements of $\beta$ in a number of nearby galaxies yield values that are on the order of $\approx 2$, but variations are observed around this value (e.g., Zhang \& Fall 1999; de Grijs \& Anders 2006; Selman \& Melnick 2008; Larsen 2009; Fall \& Chandar 2012). We adopt a fiducial value of $\beta=2$ and, in some models, we explore the effect of varying $\beta$ from $1.6$ to $2.4$. The minimum cluster mass in galaxies is poorly defined as low-mass clusters are generally faint and difficult to identify. We fix the value of $M_{cl,min}$ to 10 M$_{\odot}$. The maximum mass that a cluster can have in a galaxy is observed to depend on the SFR (e.g., Weidner et al. 2004; G\'{o}nzalez-L\'{o}pezlira et al. 2012; Schulz et al. 2015, but see also G\'{o}nzalez-L\'{o}pezlira et al. 2013 for the counter example of M51). In this work, we adopt the empirical relation between the SFR and the maximum cluster mass derived by Weidner et al. (2004), which is given by the following: 

\begin{equation}
M_{cl,max}= 8.5\times 10^{4} \times \left(\frac{{\rm SFR}}{{\rm M}_{\odot} {\rm yr^{-1}}}\right)^{0.75}.
\label{eq9}
\end{equation}

The normalization of Eq.~\ref{eq9} should also be applicable to the Milky Way. The observed Galactic SFR of $\approx 1$ M$_{\odot}$ (Robitaille et al. 2010) should be matched by a maximum mass of clusters of $\approx 8.5\times10^{4}$ M$_{\odot}$. The most massive cluster in the Galaxy is the Arches cluster and estimates of its mass fall in the range of $2\times10^{4}$ M$_{\odot}$ to $1.5\times10^{5}$ (Dib et al. 2007; Clarkson et al. 2012). A value of $8.5\times10^{4}$ appears to be an intermediate value between these estimates.  

\section{The IGIMF in the presence of cluster-to-cluster variations}\label{igimf}

\begin{figure}
\begin{center}
\includegraphics[width=0.8\columnwidth] {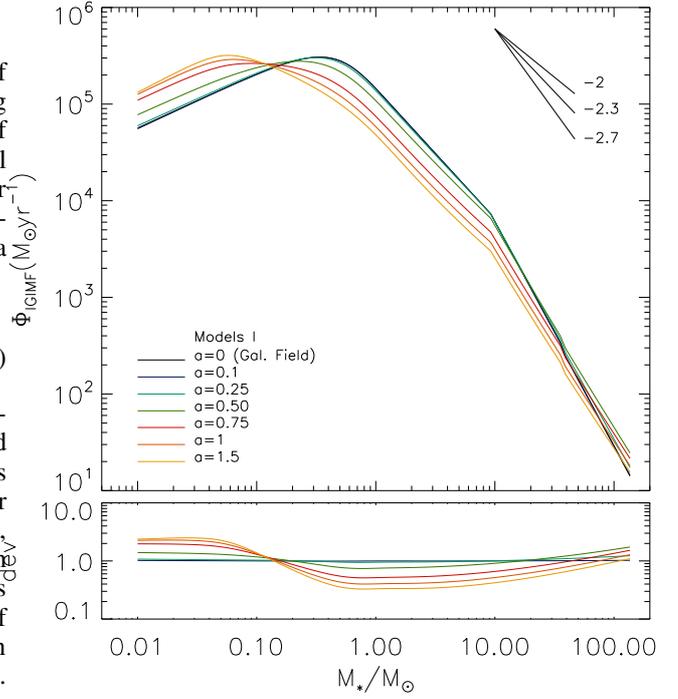}
\vspace{1cm}
\caption{IGIMF in the presence of IMF variations. Upper panel: IGIMF for the case where the IMF of clusters has the same shape ($a=0$, reference model), and for models where each of the parameters of the IMF ($\Gamma$, $\gamma$, and $M_{ch}$) has a Gaussian distribution. The width of these Gaussian distributions is described by the parameter $a$ which is the number of times the standard deviations of each of these parameters measured by Dib (2014) for young clusters in the Milky Way ($\sigma_{\Gamma}=0.6$, $\sigma_{\gamma}=0.25$, and $\sigma_{M_{ch}}=0.27$ M$_{\odot}$). The other input quantities such as the slope of the cluster mass function, minimum and maximum stellar masses, and minimum and maximum cluster mass are the same in all models (see text for details). Lower panel: The ratio of each model to the reference model.}
\label{fig1}
\end{center}
\end{figure}

\begin{figure}
\begin{center}
\includegraphics[width=0.9\columnwidth] {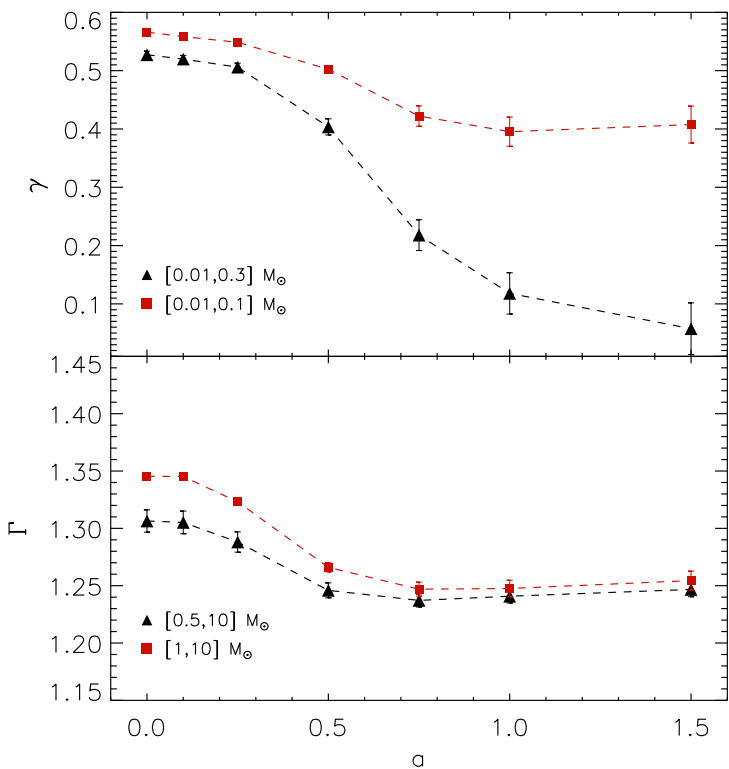}
\vspace{1cm}
\caption{Slope of the IGIMF at the low-mass end, specifically when the IGIMF was fitted with power laws in the mass ranges [0.01,0.1] M$_{\odot}$ and [0.01,0.3] M$_{\odot}$ (upper subpanel), and the slope of the IGIMF at the high-mass end when the IGIMF was fitted with a power law in the mass ranges [0.5,10] M$_{\odot}$ and [1,10] M$_{\odot}$ (lower subpanel) These calculations are based on the models presented in Fig.~\ref{fig1}.}
\label{fig2}
\end{center}
\end{figure}

In the presence of cluster-to-cluster variations of the IMF, the IGIMF, defined over a short time span $\tau_{cl}$ which is the typical lifespan of clusters in the embedded and semi-embedded phases, is the convolution of the clusters' IMF with the distribution functions of the IMF parameters, and the ICLMF, $\xi_{cl} \left(M_{cl}\right)$. When the IMF is described by the TPL function, the IGIMF is given by

\begin{eqnarray}
 \Phi_{{\rm IGIMF}}(M_{*},\Sigma_{cl},{\rm SFR},[{\rm Fe/H]})  =  \nonumber \\
   B_{*} \left(\Sigma_{cl}\right) \int_{M_{cl,min}}^{M_{cl,max(SFR)}}\int_{\Gamma_{min}}^{\Gamma_{max}} \int_{\gamma_{min}}^{\gamma_{max}} \int_{M_{ch,min}}^{M_{ch,max}} \xi_{cl}(M_{cl})\times  \nonumber \\
   \phi(M_{*},\Gamma,\gamma,M_{ch})  P(\Gamma) P(\gamma) P(M_{ch}) dM_{cl}~d\Gamma~d\gamma~dM_{ch}.
\label{eq10}
\end{eqnarray}

The normalization coefficient $B_{*}$ is simply set by the total mass available in all clusters for a given ICLMF, $\Sigma_{cl}$, with $\Sigma_{cl}=B_{*} \int_{M_{cl,min}}^{M_{cl,max}} M_{cl}^{-\beta+1} dM_{cl}$. For a given galactic $\rm SFR$ and a timescale of interest ($\tau_{cl}$), the value of $\Sigma_{cl}$ is given by $\Sigma_{cl} \approx {\rm SFR} \times \tau_{cl}$, assuming that the ${\rm SFR}$ is constant over the timescale $\tau_{cl}$. In all calculations that follow, we use $\tau_{cl}=10$ Myrs.

\subsection{Effect of the variation of the IMF parameter distributions}\label{effectvar}

In the first instance, we evaluate how the existence of cluster-to-cluster variations of the IMF affects the shape of the IGIMF. The reference model is one where the IMF of all clusters in the ICLMF is similar to the mass function of the Galactic field, with values of the three parameters being fixed to $\Gamma=1.35$, $\gamma=0.58$, and $M_{ch}=0.42$ M$_{\odot}$. Furthermore, for this model, we adopt a value of the ${\rm SFR}=1$ M$_{\odot}$ yr$^{-1}$, an exponent of the ICLMF $\beta=2$, and impose the SFR-$M_{cl,max}$ relation described by Eq.~\ref{eq9}. Here, we do not impose an  $M_{cl}-M_{*,max}$ relation and the minimum and maximum stellar masses are taken to be $M_{*,min}=0.01$ M$_{\odot}$ and $M_{*,max}=150$ M$_{\odot}$, respectively. The result for this reference model is displayed in Fig.~\ref{fig1} with the black line. Its shape is very similar to the Galactic field mass function with the exception of a steepening for masses larger than 10 M$_{\odot}$, which is a result of the choice of the minimum cluster mass.  

As discussed in Sect.~\ref{meanparam}, the variability of the IMF is included by considering that each of the parameters of the IMF for clusters found in the ICLMF belongs to a Gaussian distribution (Eqs.~\ref{eq5}-\ref{eq7}) centered around the corresponding Galactic field value. A set of values $a_{\Gamma}=1$, $a_{\gamma}=1$, and $a_{M_{ch}}=1$ (hereafter grouped with the a unique parameter $a=1$) implies that all three parameters have a standard deviation similar to those found in the Milky Way by Dib (2014) and Dib (2017). A value of $a=0.5$ implies that all three parameters have standard deviations that are half of those derived from Milky Way observations. The constraints on the scatter of these parameters were obtained using information from young clusters in the Milky Way. However, Mor et al. (2019) found that models of Dib \& Basu (2018) with $a=0.5$ best match the time-averaged IMF they inferred when comparing their chemo-dynamical models of the Galaxy with GAIA and APOGEE data. It is probably safe to say that he standard deviation for each of the IMF parameters has yet to be determined for populations of clusters that formed at the same cosmic epoch. We recall that a value of $a=1$ corresponds to the following standard deviations for the three parameters ($\sigma_{\Gamma}=0.6$, $\sigma_{\gamma}=0.25$, and $\sigma_{M_{ch}}=0.27$ M$_{\odot}$). We consider cases with $a=0.1,0.25,0.5,0.75,1,$ and $1.5$. The comparison between the IGIMF models calculated in the presence of IMF variations with the reference model (i.e., a universal IMF) is displayed in Fig.~\ref{fig1} (top subpanel). The ratio between each IGIMF with a variable IMF and the reference IGIMF is displayed in the lower subpanel of Fig.~\ref{fig1}. 

As can be observed in Fig.~\ref{fig1}, the main effect of IMF variations is to shift the peak of the IGIMF toward lower masses. As already pointed out by Dib \& Basu (2018), this effect is mostly due to variations in the characteristic mass. The slopes at the low- and high-mass ends are also shallower than the present-day mass function of the Milky Way for increasing values of $a$. Fig.~\ref{fig2} displays the derived values of the slopes in the low-mass ($\gamma$) and high-mass regimes ($\Gamma$) when the models in Fig.~\ref{fig1} are fitted with power laws in different mass ranges, namely the ranges of [0.01,0.1] M$_{\odot}$ and [0.01,0.3] M$_{\odot}$ for $\gamma$ and the mass range [0.5,10] M$_{\odot}$ and [1,10] M$_{\odot}$ for $\Gamma$. The values of both $\gamma$ and $\Gamma$ indicate shallower slopes in the low- and high-mass regime when larger variations of the IMF are included. The uncertainties on the derived slopes are larger for larger values of $a$ as the IGIMF deviates more from a single power law in the selected mass ranges. The shift in the characteristic mass of the IGIMF is larger for larger levels of IMF variations (i.e., larger values of $a$). The characteristic mass of the IGIMF is shifted from $0.42$ M$_{\odot}$ for the reference model to $\approx 0.25$ M$_{\odot}$ for $a=0.5$ and down to $\approx 0.07$ M$_{\odot}$ for $a=1$. This result offers a plausible explanation for the observed bottom-heavy stellar mass function of elliptical galaxies. As most elliptical galaxies form from the merger of disk galaxies, the physical conditions of the star-forming gas during the different phases of the merger process as well as in different locations of the merging system at any given epoch display a larger degree of variations than those that can be found for star formation occurring in a quiescent spiral galaxy (e.g., Pan et al. 2018; Bournaud et al. 2011). This large scatter in the physical conditions of the gas can lead to broad distributions of the IMF and hence to a stellar mass function that would peak at lower masses. The stellar mass function of elliptical galaxies is observed to be more top heavy with increasing stellar velocity dispersion and with enhanced $\alpha$-elements' enhancement (e.g., Conroy \& van Dokkum 2012). More detailed models that account for the variation of the IMF in clusters, the star formation history of the galaxy including its merger events, as well as for its chemical evolution are therefore needed. Van Dokkum et al. (2017b) also point out that it is the cores of the elliptical galaxies that display bottom-heavy stellar mass functions, rather than the galaxies as a whole. This picture fits with the idea presented above as it is toward the central region of the galaxy that most of the gas falls (and forms stars) during a merger event (e.g., Hernquist 1989; Barnes \& Hernquist 1991; Hopkins et al. 2008; Blumenthal \& Barnes 2018; Hani et al. 2018).

In this demonstration of the effect of IMF variations, as mean values of the Gaussian parameter distributions, we adopted  those that describe the Galactic field. As such, this demonstration is valid for star formation occurring in the present-day Milky Way. It can be argued that if this was to be true at all times, and given that the existence of IMF variations shifts the characteristic mass to lower values, it would be difficult to reconcile this result with the observed value of the characteristic mass in the Milky Way as it is observed today. However, most stars in the Galaxy formed at earlier epochs, mostly at a redshift of $\approx 2$ where the global SFR of the Galaxy was much higher, and hence the mean characteristic mass could also be higher (see the next section). The presence of IMF variations would still shift the peak of the mass function to lower masses, bringing it into agreement with the observations. The same argument is valid for elliptical galaxies whose cores were primarily assembled during intense star formation at around a redshift $\approx 2$ (Oser er al. 2010). Quantitatively, the shift to lower masses would depend on the value of the SFR and the level of IMF variations present in these systems at the time of the burst in star formation.
    
\subsection{Effect of the SFR}\label{effectsfr}

\begin{figure*}
\begin{center}
\includegraphics[width=0.9\textwidth] {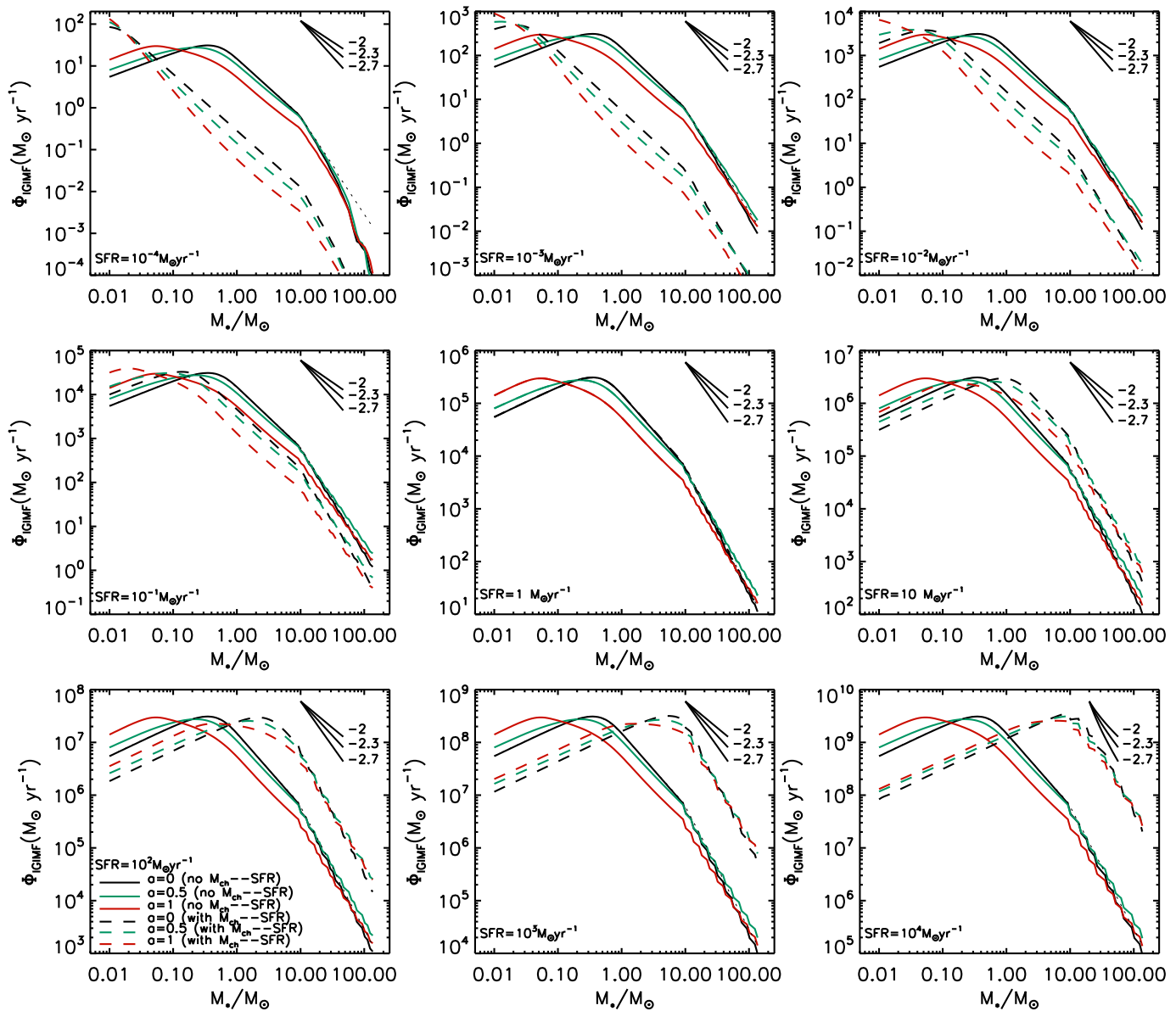}
\vspace{1cm}
\caption{Models of the IGIMF with an SFR dependence. Full lines: Models of the IGIMF with an imposed SFR-$M_{cl,max}$ relation (Eq.~\ref{eq9}). Each subpanel contains models in which the IMF is universal ($a=0$) and variable ($a=0.5$, $a=1$). Dashed lines: Similar to full line models with the additional constraint that the characteristic mass of the IMF, $M_{ch}$, is dependent on the SFR (Eq.~\ref{eq4}).}
\label{fig3}
\end{center}
\end{figure*}

\begin{figure}
\begin{center}
\includegraphics[width=0.9\columnwidth] {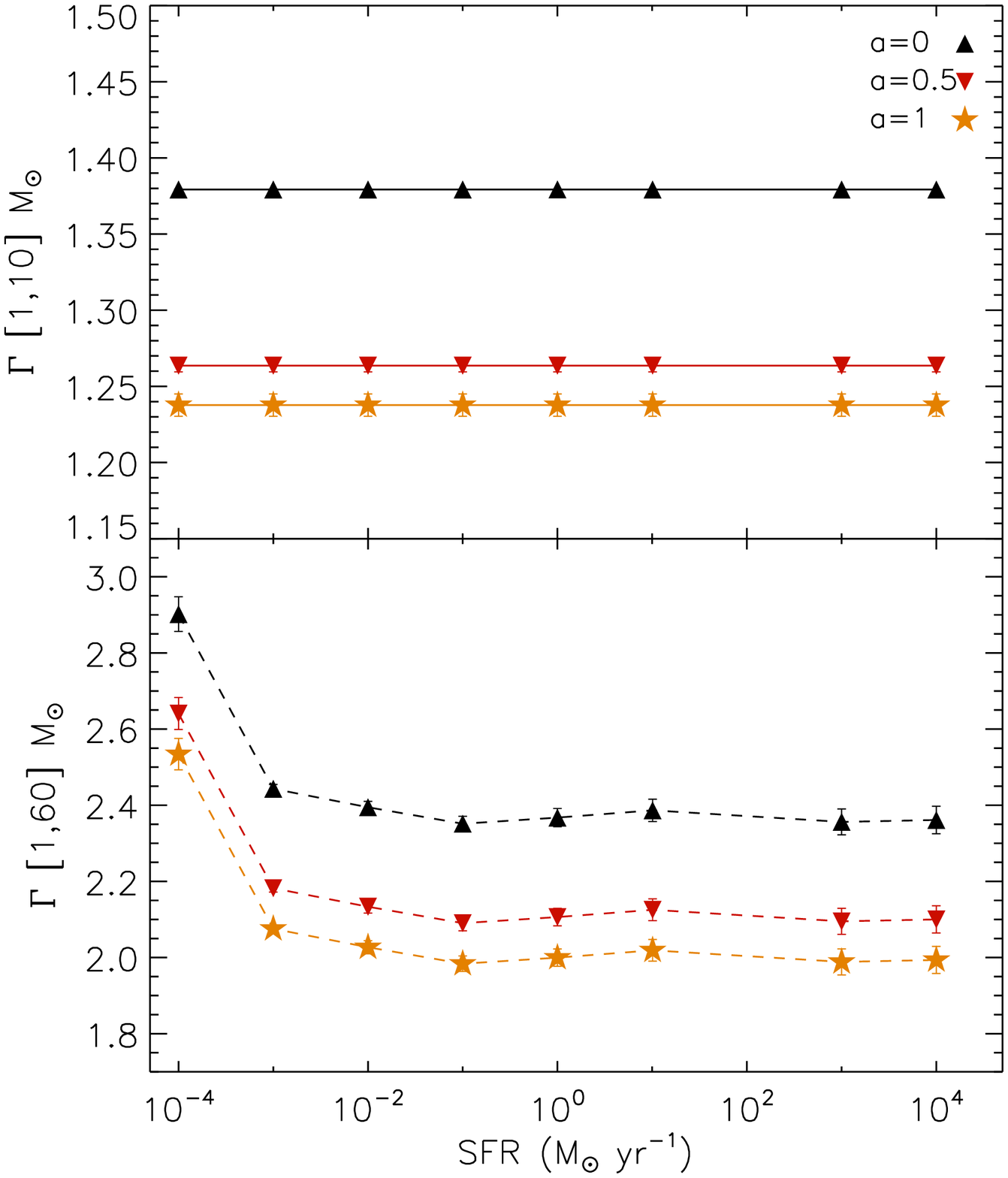}
\vspace{1cm}
\caption{Dependence of the slope of the IGIMF in the high-mass end on the SFR. The upper subpanel displays the slope of the IGIMF in the mass range $[1,10]$ M$_{\odot}$ as a function of the SFR for three values of the IMF variation parameter $a$. The lower subpanel displays the slope of the IGIMF in the mass range [1-60] M$_{\odot}$ as a function of the SFR and similarly for the values of $a=0, 0.5$ and $1$. These calculations are based on the models displayed in Fig.~\ref{fig3} which have an implemented $M_{cl,max}-$SFR relation, but without the $M_{ch}$-SFR dependence.}
\label{fig4}
\end{center}
\end{figure}

In the model presented here, the shape of the IGIMF can be affected by the SFR in two different ways. The first of these two effects is related to the existence of the empiriral SFR-$M_{cl,max}$ relation (Eq.~\ref{eq9}). For a fixed shape of the ICLMF, this relation implies that galaxies with a high SFR harbor higher mass clusters, and this could affect the IGIMF at the high-mass end. The second effect depends on the validity of the SFR-$M_{ch}$ conjecture and this would result in a higher characteristic mass at higher SFRs. Figure \ref{fig3} displays calculations of the IGIMF for various values of the galactic SFR starting from $10^{-4}$ M$_{\odot}$ yr$^{-1}$ and up to $10^{4}$ M$_{\odot}$ yr$^{-1}$. A version of these models implements only the SFR-$M_{cl,max}$ relation (full lines), and a second one implements both the SFR-$M_{cl,max}$ and the SFR-$M_{ch}$ relations (dashed lines). In all of these models, we fix the exponent of the ICLMF to $\beta=2$. For each of these models, we consider cases where cluster-to-cluster variations are absent and where the IMF of all clusters is universal and considered to be the same as the present-day mass function of the Milky Way ($a=0)$, and other cases where IMF variations are moderate ($a=0.5$) and large ($a=1$). 

For models that implement only the SFR-$M_{cl,max}$ relation, the sole effect of an increase in the SFR is to flatten the slope of the IGIMF at the high-mass end moderately. Figure~\ref{fig4} displays the slope of the IGIMF in the mass ranges of [1,10] M$_{\odot}$ (upper subpanel) and [10,60] M$_{\odot}$ (lower subpanel) as a function of the SFR for the three values of $a$. In the intermediate stellar mass range [1,10] M$_{\odot}$, the slope of the IGIMF is independent of the SFR. However, at the high-mass end, the slope of the IGIMF is steeper with decreasing SFR. Figure~\ref{fig4} shows that with the inclusion of the SFR-$M_{cl,max}$ relation, the IGIMF variations remain dominated by IMF variations. On the other hand, an SFR-dependent characteristic mass significantly modifies the IGIMF. At low SFRs, the IGIMF nearly resembles a single power-law function and is steeper when IMF variations are larger. This is due to the fact that both a lower SFR and a smaller level of IMF variations jointly contribute to pushing the characteristic mass to lower values. At increasing SFRs, the effect of IMF variations is more noticeable around the characteristic mass. Larger variations (i.e., larger values of $a$) lead to broader peaks of the IGIMF with the previously discussed effect of the characteristic mass being shifted to lower values in the presence of larger IMF variations. The gradual global shift to higher characteristic masses with an increasing SFR leads to steeper slopes at the high-mass end. 

\subsection{Effect of the initial cluster mass function}\label{effectclmf}

\begin{figure*}
\begin{center}
\includegraphics[width=0.9\textwidth] {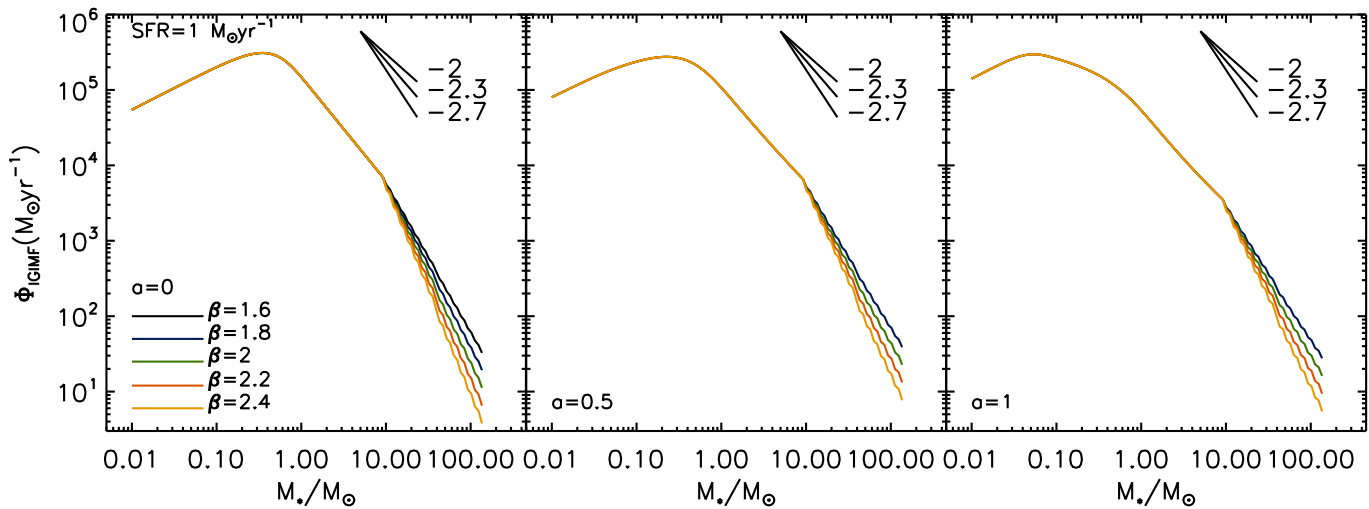}
\vspace{1cm}
\caption{Models of the IGIMF for the case SFR=1 M$_{\odot}$ yr$^{-1}$ and for various values of the exponent of the ICLMF. Smaller values of $\beta$ correspond to a shallower ICLMF. The three subpanels correspond to cases with $a=0$ (left subpanel), $a=0.5$ (mid subpanel), and $a=1$ (right subpanel).}
\label{fig5}
\end{center}
\end{figure*}

In all previous calculations, we adopted a value of $\beta=2$ for the exponent of the ICLMF. Here, we explore the effect of varying the value in $\beta$ in conjunction with variations in the IMF. We consider values of $\beta=1.6, 1.8, 2.2,$ and $2.4$ along with the fiducial value of $\beta=2$. In all models, we consider a value of ${\rm SFR}=1$ M$_{\odot}$ yr$^{-1}$. For each value of $\beta$, we calculate three models, a reference model with $a=0$ (i.e., no IMF variations) and two other cases with $a=0.5$ and $a=1$. The models with different values of $\beta$ are displayed in Fig.~\ref{fig5}: the reference models are displayed in the left subpanel and the models with $a=0.5$ and $a=1$ are displayed in the middle and right panel, respectively. As can be seen in Fig.~\ref{fig5}, the value of $\beta$ affects the slope of the IGIMF at the high-mass end. Smaller values of $\beta$ correspond to a shallower ICLMF which contains a relatively larger fraction of massive clusters, and hence larger numbers of massive stars, leading to a shallower IGIMF. However, the shape of the IGIMF is largely unaffected by the value of $\beta$ in the intermediate and low stellar mass regimes and this is valid irrespective of the level of IMF variations. 
 
 \subsection{Effect of the $M_{cl}-M_{*,max}$ relation}

The existence of the $M_{cl}-M_{*,max}$ relation is still debated. On the one hand, there are groups who have argued that this relation is deterministic (e.g., Vanbeveren 1982; Weidner \& Kroupa 2004; Weidner et al. 2013a), while other interpretations argue that the relation is a natural consequence of stochastic sampling of the IMF in clusters of different masses (e.g., Calzetti et al. 2010; Hermanowicz et al. 2013; Popescu \& Hanson 2014). Dib et al. (2017) found that imposing an $M_{cl}-M_{*,max}$ relation inhibits the reproduction of the fraction of single O stars in a subsample of Galactic clusters. It is useful to point out that including the $M_{cl}-M_{*,max}$ relation can only modify the IGIMF at the high-mass end. We verified how the $M_{cl}-M_{*,max}$ relation modifies the IGIMF with and without IMF variations. The latest version of the $M_{cl}-M_{*,max}$ relation is given by ${\rm log}_{10}(M_{*,max}/{\rm M}_{\odot})=-0.66+1.08\times \left[{\rm log}_{10}(M_{cl}/{\rm M}_{\odot})\right]-0.15\times\left[{\rm log}_{10}(M_{cl}/{\rm M}_{\odot})\right]^{2}+0.0084\times\left[{\rm log}_{10}(M_{cl}/{\rm M}_{\odot})\right]^{3}$ (Weidner et al. 2013a) and it is assumed to be valid for cluster masses $M_{cl} \lesssim 2.5 \times 10^{5}$ M$_{\odot}$. 

\begin{figure*}
\begin{center}
\includegraphics[width=0.9\textwidth] {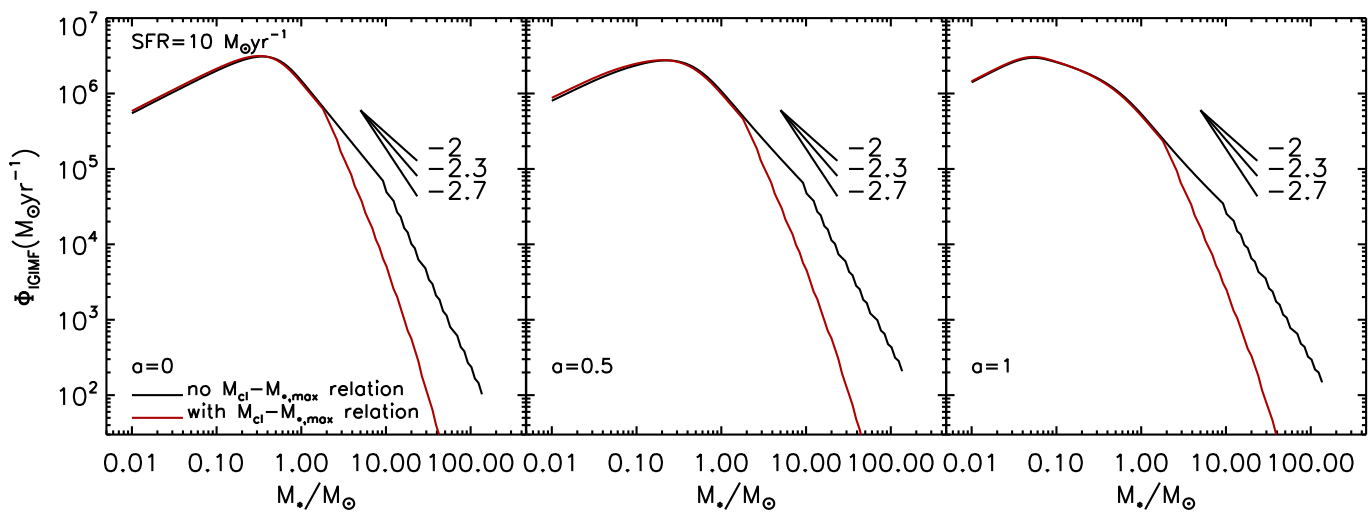}
\vspace{1cm}
\caption{Models of the IGIMF for an SFR=10 M$_{\odot}$ yr$^{-1}$ with and without the $M_{cl}-M_{*,max}$ relation. The cases correspond to $a=0$ (left subpanel), $a=0.5$ (mid subpanel), and $a=1$ (right subpanel).}
\label{fig6}
\end{center}
\end{figure*}

We calculated a set of models with a fixed value of the SFR=10 M$_{\odot}$ yr$^{-1}$ and $\beta=2$. As in all other models, the SFR-$M_{cl,max}$ relation was also accounted for. We considered cases with $a=0$, $a=0.5,$ and $a=1$. The comparison of the IGIMF is displayed in Fig.~\ref{fig6} for the cases with $a=0$ (left subpanel), $a=0.5$ (mid subpanel), and $a=1$ (right subpanel). As can be seen in Fig.~\ref{fig6}, the inclusion of the $M_{cl}-M_{*,max}$ relation has no influence on the shape of the IGIMF in the low- to intermediate-mass regime where only the presence of IMF variations plays a role in shifting the characteristic mass of the IGIMF to lower masses. However, in the regime of high stellar masses, the effect of the $M_{cl}-M_{*,max}$ relation dominates over the effect of IMF variations rendering the IGIMF steeper, with no noticeable effects due to IMF variations.  

\subsection{Effect of density and metallicity}\label{metaldens}

\begin{figure*}
\begin{center}
\includegraphics[width=0.9\textwidth] {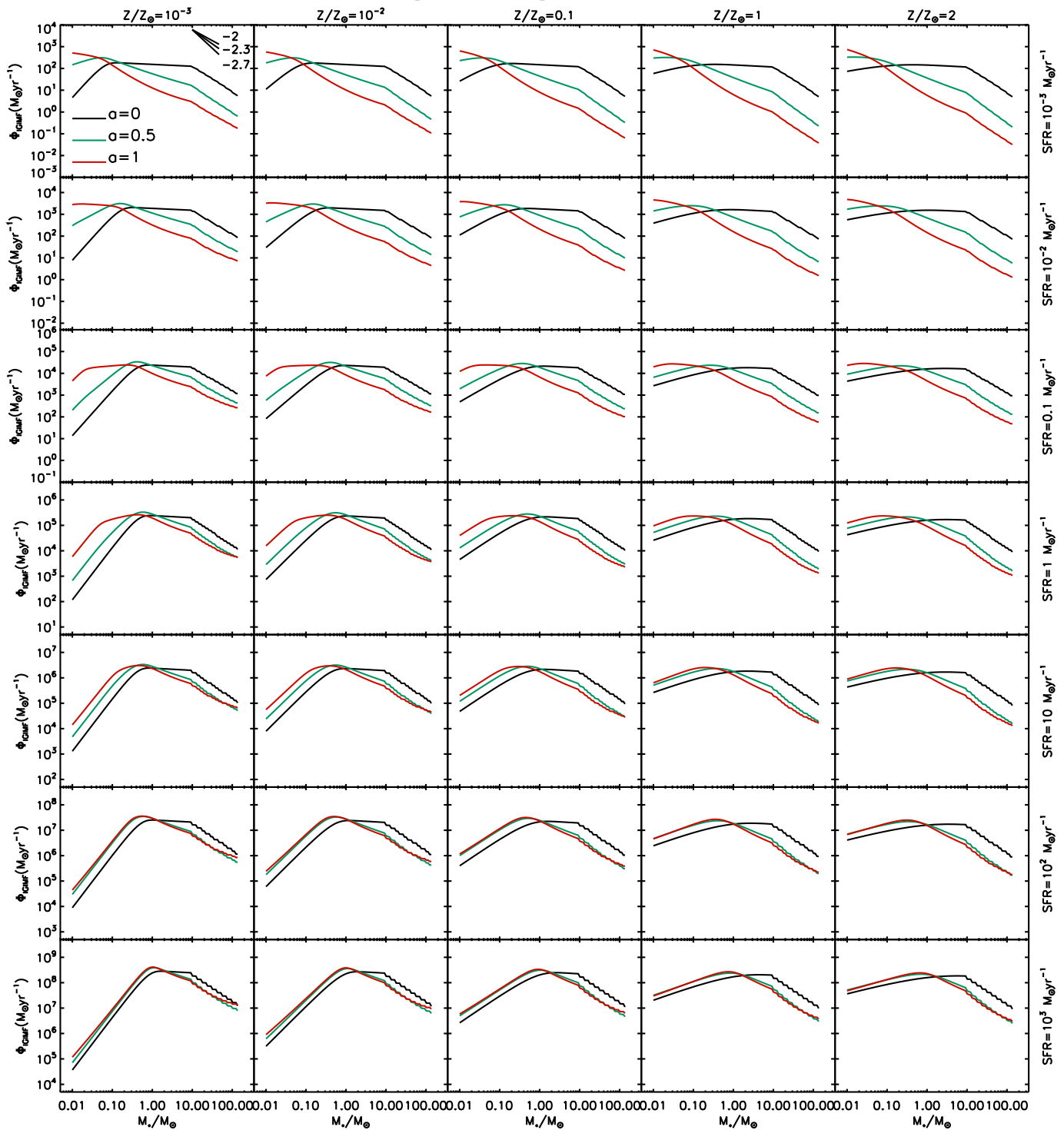}
\vspace{1cm}
\caption{Models of the IGIMF for various values of the SFR, ranging from $10^{-3}$ M$_{\odot}$ yr$^{-1}$ to $10^{3}$ M$_{\odot}$ yr$^{-1}$ (rows), and metallicity, ranging from $Z/Z_{\odot}=10^{-3}$ to 2. (columns). Each subpanel displays the case with $a=0$ (black line), $a=0.5$ (green line), and $a=1$ (red line).}
\label{fig7}
\end{center}
\end{figure*}

\begin{figure*}
\begin{center}
\includegraphics[width=0.9\textwidth] {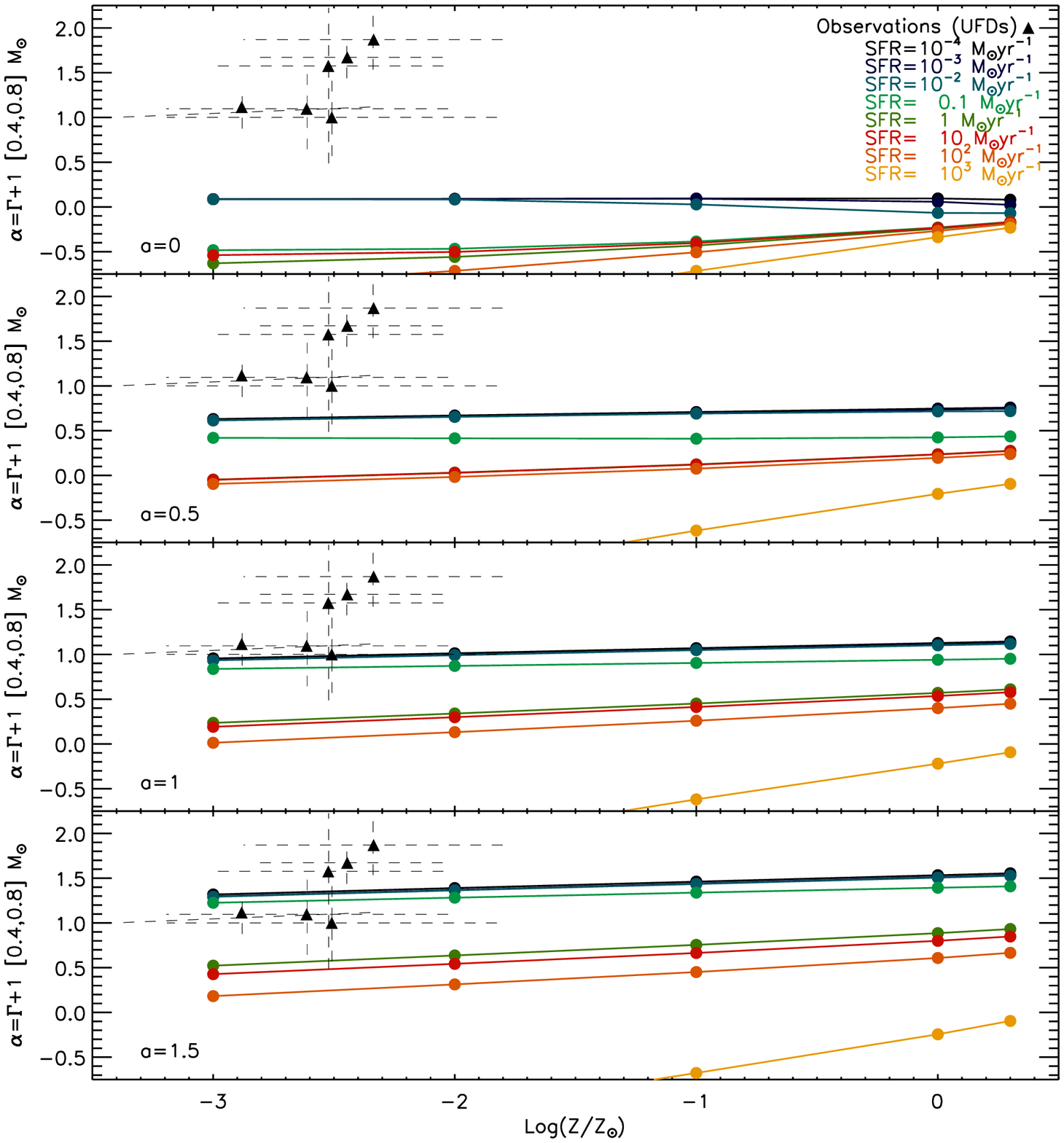}
\vspace{1cm}
\caption{Slope of the IGIMF in the stellar mass range [0.4,0.8] M$_{\odot}$ for all models displayed in Fig.~\ref{fig7}. Additional models with an SFR$=10^{-4}$ M$_{\odot}$ yr$^{-1}$ as well as with $a=1.5$ are included. The observational data points correspond to six UFD satellites of the Milky Way and are taken from Gennaro et al. (2018).}
\label{fig8}
\end{center}
\end{figure*}

As discussed in Sect.~\ref{mathimf}, Marks et al. (2012) propose, based on observational data, that the high-mass end of the IMF in stellar clusters depends on the cluster metallicity and total initial density (Eqs. \ref{eq2} and \ref{eq3}). The dependence on density can be transformed into a dependence on cluster mass under specific assumptions. The cluster density can be written as $\rho_{cl}=\left(3 M_{cl}/4\pi r_{h}^{3}\right)$, where $r_{h}$ is the cluster half-mass radius. Assuming a fixed star formation efficiency that is independent of the protocluster mass, one can write that $\rho_{pcl}=3 M_{pcl}/4\pi r_{h}^{3}$. Marks \& Kroupa (2012) found that the half-mass radius scales as $r_{h}/{\rm pc}=0.1 (M_{cl}/{\rm M_{\odot}})^{0.13}$ and also that ${\rm log_{10}} \rho_{cl}=0.61 {\rm log_{10}}M_{cl}+2.08$. With these scalings, it is possible to rewrite Eq.~\ref{eq3} as follows: 

\begin{equation}
x=-0.14[{\rm Fe/H}]+0.6 {\rm log}_{10} \left(  \frac{M_{cl}} {10^{6}{\rm M_{\odot}}} \right)+2.83
\label{eq11}
.\end{equation}

In this work, we employ Eq.~\ref{eq11}. Another formulation for the dependence of the high-mass slope of the galactic present-day mass function on metallicity has been presented by Mart\'{i}n-Navarro (2015) and used by Gutcke \& Springel (2019) to investigate the role of a metallicity-dependent IMF on the dynamics and the chemical enrichment of the gas in galactic disks similar to the Milky Way. Their models with a metallicity-dependent IMF show a better agreement with the APOGEE data (i.e., the [$\alpha/{\rm H}$] vs. [Fe/H] relation; Bovy et al. 2014, Ness et al. (2018)) than in the case of the Galactic field-like IMF. Kroupa (2002) and Marks et al. (2012) also suggested a possible dependence of the shape of the IMF in the low-mass regime ($M_{*} < 0.5$ M$_{\odot}$) on metallicity. A multicomponent mass function, such as the one proposed by Kroupa, has two power laws in the mass regime below $0.5$ M$_{\odot}$. Translating this into a TPL mass function such as the one used in this work, one can still write this relation as follows:

\begin{equation}
\gamma=\gamma_{field}-0.5[{\rm Fe/H}]=0.58-0.5[{\rm Fe/H}]
\label{eq12}
.\end{equation}

Metallicity could be one of the physical causes for IMF variations we are considering in this work. Recent theoretical work by Sharda \& Krumholz (2022) suggests that metallicity-induced variations of the characteristic mass can be significant, especially at metallicities lower than 0.1 $Z_{\odot}$. The metallicity-dependent variations discussed above connect the metallicity to the shape of the IMF independently of other physical variables that are known to affect the star formation process, such as the nature and compression levels of turbulence and the strength of the magnetic field. Unless all other physical processes conspire and lead to the observed empirical dependence of the high-mass slope of the IMF on metallicity, it is reasonable to think that variations of the IMF would still be observed for any value of the metallicity in a given galactic population of star-forming regions. This is the case in the models of Sharda \& Krumholz (2022), and it is also the case for present-day star formation in the Milky Way where stars are forming in different star-forming regions with roughly the same metallicity and where variations of the IMF are observed (e.g. Dib 2014; Dib et al. 2017).

In galaxies as a whole, there is likely a causal relationship between the galaxy's star formation history, the time-dependent IGIMF, and the galaxy's averaged gas-phase metallicity (e.g., Sanch\'{e}z-Menguiano et al. 2019). Here, we have taken a simpler approach and calculated the IGIMF over a grid of distinct SFRs and metallicities. Combining the metallicity effects described in Eq.~\ref{eq11} and Eq.~\ref{eq12}, along with the density dependence, we constructed a set of IGIMF models for various values of the SFR and metallicity with and without IMF variations. As before, we considered three values of $a$, namely $a=0$ (no IMF variations), $a=0.5$, and $a=1$. We considered SFRs in the range $10^{-3}$ M$_\odot$ yr$^{-1}$ and $10^{3}$ M$_{\odot}$ yr$^{-1}$ and values of [Fe/H] that span the range $-3$ to $0.3$, corresponding to $Z/Z_{\odot}$ from $\approx 10^{-3}$ to $2$. All models have a value of $\beta=2$ for the exponent of the ICLMF, and implement the SFR-$M_{cl,max}$ relation, but not the hypothetical SFR-$M_{ch}$ dependence. 

The results of these calculations are displayed in Fig.~\ref{fig7}. Each subpanel of Fig.~\ref{fig7} shows cases with the three considered values of $a$, namely $a=0$ (black lines), $a=0.5$ (green lines), and $a=1$ (red lines). Columns correspond to the same value of metallicity and rows correspond to the same value of the SFR. At low metallicities and low SFR, the IGIMF displays a power-law-like behavior and its steepness increases in the presence of increasingly significant IMF variations. These results are compatible with those observed for low surface brightness, metal-poor galaxies (Lee 2004; Hoversten \& Glazebrook 2008). For a fixed value of the SFR, the IGIMF is steeper at lower metallicities (i.e., Eq.~\ref{eq12}) and shallower in the high-mass regime. These results are compatible with the findings of Mart\'{i}n-Navarro et al. (2015) which argue for a shallower slope at the high-mass end with decreasing metallicity. In the presence of IMF variations, the IGIMF transitions from a single power-law-like function to a composite power law at an SFR of $\approx 1$ M$_{\odot}$. 
 
 \section{Comparison to the observations of ultra-faint dwarf galaxies}\label{compobs}

It is important to compare our results to different subsets of galaxies for which measurements of the shape of the IGIMF are available. A direct comparison to the present-day stellar mass function of both late-type and early-type galaxies is problematic given that the present-day mass function is the integrated IGIMF of the galaxy over cosmic time. However, an interesting case is that of ultra-faint dwarf (UFD) galaxies. In these galaxies, where star formation occurred early at around a redshift $\approx 6$ and has been subsequently quenched (e.g., Sacchi et al. 2021), the present-day stellar mass function of the galaxy for stars whose lives on the main sequence are longer than the age of these galaxies (i.e., low-mass stars) is a good representation of the IGIMF when most stars have formed. Furthermore, the stellar mass function of UFD galaxies is unlikely to be affected by dynamical evolution. An estimate of the relaxation time for such galaxies is $\approx 10^{4}$ Gyr, which is in significant excess of the age of the Universe (Gennaro et al. 2018). Gennaro et al. (2018) derived the metallicity and the galactic stellar mass function for six satellite UFD galaxies of the Milky Way. By assuming a single power-law shape, they measured the slope of the stellar mass function over the stellar mass range [$0.4,0.8$] $\odot$. 

The metallicities of the six UFD galaxies fall in the range $Z/Z_{\odot} \approx 10^{-3}$ to $10^{-2}$. In order to enable the comparison with our models, we computed the slope of the IGIMF in the stellar mass range [0.4,0.8] M$_{\odot}$ for all models displayed in Fig.~\ref{fig7}. We added models with SFR$=10^{-4}$ M$_{\odot}$ yr$^{-1}$ as well as models with $a=1.5$. The results are displayed in Fig.~\ref{fig8} for four values of $a$, ranging between $0$ and $1.5$. The comparison of the models to the observations clearly shows that a good agreement with the observational data is achieved for models with $a=1$ and $a=1.5$. We recall here that $a=1$ corresponds to standard deviations of the parameters of the IMF that are of the same order as those derived for a population of young clusters in the Milky Way (Dib 2014; Dib et al. 2017) and they correspond to $\sigma_{\Gamma}\approx 0.6$, $\sigma_{M_{ch}} \approx 0.27$ M$_{\odot}$, and $\sigma_{\gamma}=0.25$. The same comparison to the Gennaro et al. (2018) data has been attempted by Je\v{r}\'{a}bkov\'{a} et al. (2018). However, their models of the IGIMF, which do not included IMF variations in clusters, failed to reproduce the observations. A reassuring aspect is that their models that are the closest to our results for the universal IMF case with a metallicity and density dependence show the same trend as what is observed in the top subpanel of Fig.\ \ref{fig8} (their model IGIMF3 vs. the case with $a=0$ in Fig.~\ref{fig8}).

Discussing the physical origin of all possible causes of IMF variations is beyond the scope of this work, simply because the literature is too extensive to quote is a single discussion section. In the different sections above, we have discussed some theoretical results that suggest whether the shape of the IMF varies with the environment (e.g., Dib et al. 2007,2010, Hocuk et al. 2010,2011, Sharda \& Krumholz 2021). Yet, it is useful to point to a few additional works that argue in favor of the existence of local variations of the IMF in galaxies. For example, Ballero et al. (2013) tested whether a universal IMF (i.e., in their case, the functional form suggested by Kroupa 2001) is sufficient to explain the metallicity distributions observed in the bulge of the Milky Way and in M31. Their conclusion is that some variations of the IMF around a value that is shallower than the Salpeter slope (i.e., $\Gamma \approx 1$) are necessary in order to reproduce the observations. Bekki \& Meurer (2013) took a different approach in which they implemented the metallicity and density-dependent IMF prescribed by Marks et al. (2012) (E.~\ref{eq3}). Using this prescription of the IMF where implicitly IMF variations are driven by the gas-phase metallicity of the gas and by its density, they were able to recover the observed dependence of the slope of the IMF in the high stellar mass regime on the SFR (Gunawardhana et al. 2011). {Deviations from a Salpeter slope at the high-mass have also been reported in high stellar surface density clusters (e.g., Lu et al. 2013, Hosek et al. 2019) as well as in the cores of elliptical galaxies (van Dokkum et al. 2017b)}. More recently, Dib et al. (2022) have shown that allowing for variations of the IMFs, particularly in the regime of stellar masses around its peak, can help explain the scatter that is observed between the fraction of first-generation stars and the present-day slope of the mass function in globular clusters that has been recently presented by Kravtsov et al. (2022). It is probably irrefutable that more work is still needed in order to better quantify the level of IMF variations within galaxies, and to relate these variations to the physical conditions of the star-forming gas.  

\section{Conclusions}\label{conc}

We explored how the IGIMF is affected by the presence of cluster-to-cluster IMF variations in a population of coeval young clusters (i.e., with an age spread between clusters of $\lesssim 10$ Myr). IMF variations are taken into account in the form of Gaussian distribution functions for each of the parameters that describe its shape. For the IMF functional form adopted in this work (i.e., a tapered power-law function), the three parameters are the slope at the high mass end, $\Gamma$, the slope at the low-mass end $\gamma$, and the characteristic mass $M_{ch}$. A larger level of IMF variations corresponds to a larger width of the parameter distributions. We have shown that when variations of the IMF are significant, the main impact on the IGIMF is to shift its peak toward lower masses, in addition to moderately flattening the slope of the IGIMF at the low- and high-mass ends. The shift to lower masses is more prominent when IMF variations are large. IMF variations are readily observable in the Milky Way and they span the entire spectrum of stellar masses. Such variations could be even larger in interacting and merging galaxies since the physical conditions (temperature, density, radiation, and external pressure) can vary greatly during and across the merging system (e.g., Bournaud et al. 2011; Pan et al. 2018). Large IMF variations lead to an IGIMF that resembles a single power-law function. This could explain the observed stellar mass functions of early-type galaxies as the latter are believed to have formed from the merger of disk galaxies where the star-forming gas exhibits a large range of physical conditions. 

We also explored how IMF variations couple to other galactic properties, such as the SFR, the shape of the ICLMF, the gas-phase metallicity, and the constraint imposed by the existence of a relation between the mean characteristic mass of the IMF and the SFR (SFR-$M_{ch}$ relation) as well as the existence of a relation between the maximum stellar mass that can be found in a cluster and the cluster mass (the $M_{*,max}-M_{cl}$ relation). All models implement the observed correlation between the SFR and the maximum cluster mass (i.e., the SFR-$M_{cl,max}$ relation). Both the shape of the ICLMF and the $M_{*,max}-M_{cl}$ relation affect the shape of the IGIMF in the regime of high stellar masses and they tend to dominate over the effect of IMF variations in this stellar mass regime. Of the other effects that can modify the shape of the IGIMF, the occurrence of an SFR-$M_{ch}$ relation in clusters can generate bottom-heavy IGIMFs at low SFRs, and the IGIMF is steeper for larger variations of the IMF. At intermediate SFR values ($\approx 0.1-10$ M$_{\odot}$ yr$^{-1}$), the effects of IMF variations are as important as the effects of the SFR-$M_{ch}$ relation. While the effect of larger values of the SFR is to shift the characteristic mass to larger values, the effect of large IMF variations is to shift the characteristic mass to lower values and to generate a shallower slope at the high-mass end. For large values of the SFR ($\gtrsim 10$ M$_{\odot}$ yr$^{-1}$), the characteristic mass of the IGIMF can shift to values in excess of 1 M$_{\odot}$ and the characteristic mass of the IGIMF becomes relatively insensitive to the presence of IMF variations. We have also explored the coupling between the existence of IMF variations and metallicity and cluster density which are based on empirical observational evidence. For a fixed value of the SFR, we find that the IGIMF is shallower with decreasing metallicity at the high-mass end while staying steeper for increasing levels of IMF variations. In the presence of a metallicity dependence, the IGIMG at a low SFR ($\lesssim 0.1$ M$_{\odot}$ yr$^{-1}$) is close to a single power-law form, and it transforms gradually to a bell-like function with increasing SFR.

We compare the SFR+metallicity-dependent models of the IGIMF to the stellar mass function of UFD Milky Way satellites. As stars in these UFD galaxies are believed to have formed in a single burst at redshift $\approx 6$ and should remain relatively unaffected by subsequent dynamical evolution, their observed stellar mass functions are a good representation of the IGIMF when the majority of their stellar populations have formed at low metallicities and low SFRs. The comparison in Fig.~\ref{fig8} strongly suggests that only models that account for IMF variations coupled to a metallicity and SFR dependence can help reproduce the observations. The level of variations that are necessary to fit the observations are on the order of 1 to 1.5 times those observed for young clusters in the Milky Way.  

The results presented in this paper show that complex features can emerge in the IGIMF when IMF variations are taken into account, particularly when they are coupled to variations in the mean values of the parameters that describe the IMF, such as the mean characteristic mass and its dependence on the SFR, which may vary as a function of time. Our results suggest that the inclusion of IMF variations in models of galaxy formation and evolution is essential in order to improve our understanding of star formation and star formation feedback effects on galactic scales.

\begin{acknowledgements}
I am very grateful to the anonymous referee for constructive comments. I thank Mario Gennaro for sharing the data points on the UFD galaxies and for useful discussions. I also thank Zhiqiang Yan and Pavel Kroupa for sharing very useful comments on the first version of the paper.
 \end{acknowledgements}

{}

\label{lastpage}

\end{document}